\documentclass[%
reprint,
amsmath,
amssymb,
aps,
floatfix,
]{revtex4-2}

\usepackage{graphicx}
\usepackage{dcolumn}
\usepackage{bm}

\begin{document}

\title{Long-term electrical characteristics \\of a poly-3-hexylthiophene water-gated thin-film transistor}

\author{Axel Luukkonen}
\author{Amit Tewari}
\author{Kim Bj\"{o}rkstr\"{o}m}
\author{Amir Mohammad Ghafari}
\author{Ronald \"{O}sterbacka}
 \email{ronald.osterbacka@abo.fi}
\affiliation{
Physics, Faculty of Science and Engineering, \r{A}bo Akademi University, Henriksgatan 2, 20500 Turku, Finland
}

\author{Eleonora Macchia}
 \affiliation{Physics, Faculty of Science and Engineering, \r{A}bo Akademi University, Henriksgatan 2, 20500 Turku, Finland\\
 Dipartimento di Farmacia - Universit\`{a} degli Studi di Bari Aldo Moro, Via Orabona 4, 70125 Bari, Italy}
 
\author{Fabrizio Torricelli}
\affiliation{
Department of Information Engineering, University of Brescia, Via Branze 38, 25123 Brescia, Italy
}

\author{Luisa Torsi}
\affiliation{
Dipartimento di Chimica - Universit\`{a} degli Studi di Bari Aldo Moro, Via Orabona 4, 70125 Bari, Italy
}

\date{\today}

\begin{abstract}
Organic water-gated thin-film transistors (WG-TFTs) are of great interest in developing low-cost and high-performance biosensors. The device’s sensitivity to changes in measurement conditions can impair long-term operation, and care must be taken to ensure that the WG-TFT sensor response is due to an actual biorecognition event occurring on the sensing electrode. This work aims to clarify the long-term stability of a poly-3-hexylthiophene (P3HT) WG-TFT operated intermittently over two months during 5750 measurement cycles. We have evaluated the device figures of merit (FOM), such as threshold voltage, mobility, and trap density, during the whole measurement period. Short-term changes in the FOM are mainly attributed to work function changes on the gate electrode, whereas long-term changes are consistent with an increase in the semiconductor trap density. The shift in threshold voltage and decrease in mobility are found to be linear as a function of measurement cycles and caused by electrical stress, with time immersed in water having a negligible effect on the device. The trap density-of-states estimated using the subthreshold slope is similar to earlier reported values for P3HT OFETs and exhibits a gradual increase during device use and a partial recovery after rest, indicating the formation of shorter- and longer-lived traps. 
\end{abstract}

\maketitle

\section{\label{sec:intro}Introduction}
\begin{figure}[b]
\includegraphics[width=\linewidth]{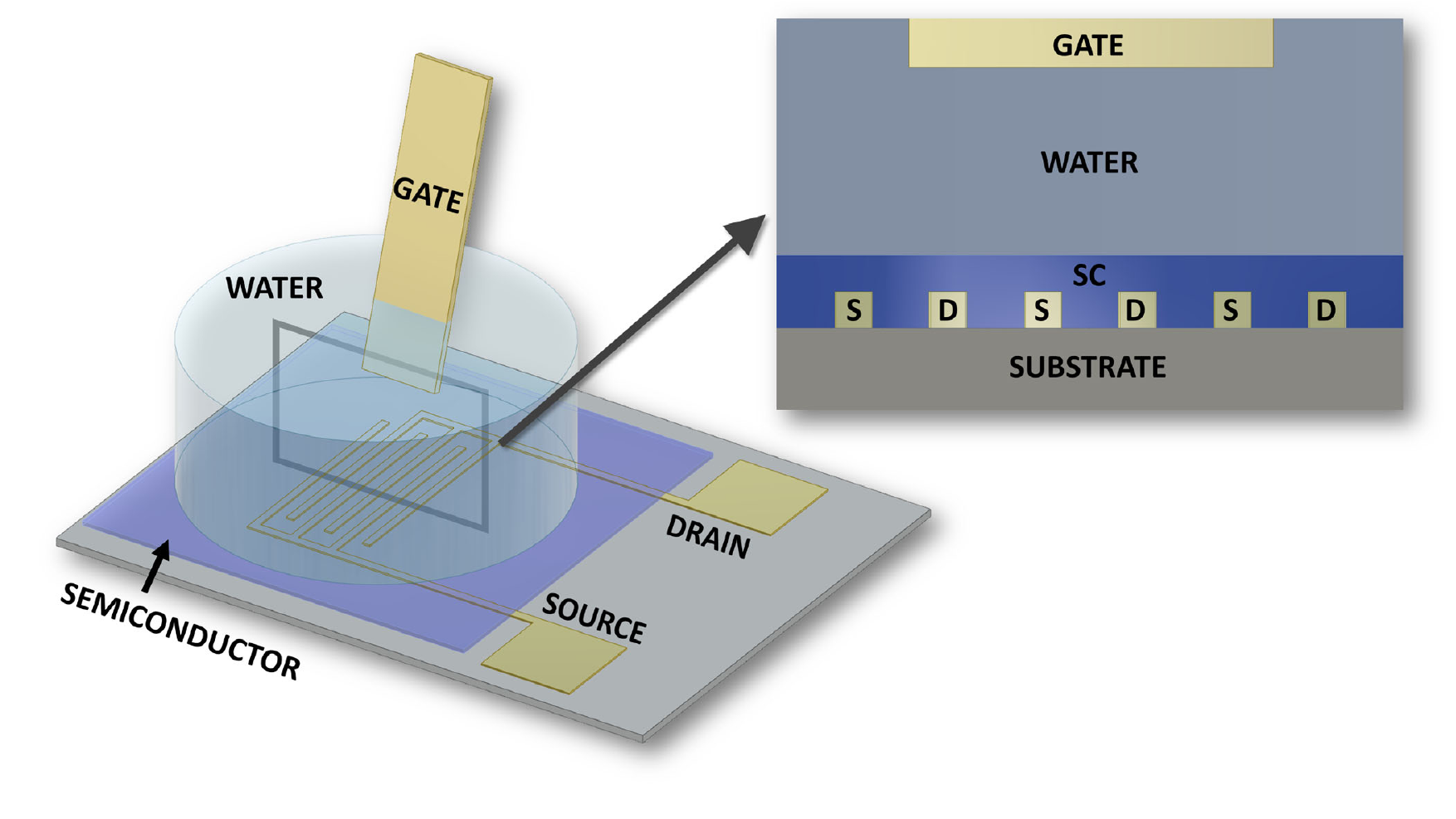}
\caption{\label{fig:1} A simple schematic of a WG-TFT with interdigitated source and drain (S-D) electrodes.}
\end{figure}

Water-gated thin-film transistors (WG-TFTs) employ water as a gate dielectric as opposed to the solid dielectric, typically $\text{SiO}_2$, used in conventional organic field-effect transistors (OFETs).\cite{kergoat_water-gate_2010,bandiello_aqueous_2014,zhang_high_2016} The WG-TFT, illustrated in Fig.~\ref{fig:1}, is made up of three parts: (i) the semiconductor deposited on the substrate and contacted with the source and drain electrodes, (ii) the aqueous gate dielectric, and (iii) the gate electrode, usually metallic. Due to the formation of electrical double layers at the gate electrode and semiconductor surface upon applying a gate voltage, the gating capacitance is considerably higher than in conventional OFETs.\cite{kergoat_water-gate_2010,porrazzo_field-effect_2015} This enables low-voltage ($<1$\,V) operation and makes the device highly sensitive to electrical or capacitive changes at the gate-electrolyte and semiconductor-electrolyte interfaces, making the WG-TFT attractive for use in biosensors.\cite{wang_electrolytic_2016,picca_ultimately_2020} While WG-TFTs have demonstrated high performance in a biosensor setting \cite{picca_ultimately_2020,ricciardi_immunodetection_2013,berto_egofet_2018,macchia_single-molecule_2018,picca_study_2019}, the device stability can still be improved to shorten measurement times and lengthen the usable device lifetime. The device should also be as stable as possible to avoid incorrect readings during sensing measurements. 

Device degradation in conventional organic field-effect transistors (OFETs) typically manifests as decreased source-drain current, increased hysteresis, a shift in threshold voltage, and increased gate leakage.\cite{gomes_electrical_2005,sirringhaus_reliability_2009,bobbert_operational_2012} An increase in trap density is the main cause of these effects.\cite{nikolka_high_2017,haneef_charge_2020} Electrical degradation in WG-TFTs is expressed similarly.\cite{kergoat_water-gate_2010,picca_study_2019,kim_electrolyte-gated_2013,porrazzo_improving_2014} However, the degradation mechanisms in WG-TFTs are challenging to pinpoint and model due to the complexity of the double layers. Furthermore, operation in aqueous environments can lead to unwanted electrochemical effects, causing increased doping and oxidation of the semiconductor and rapid degradation of the WG-TFT. These effects can be largely mitigated by operating the device within the proper voltage window and by using a hydrophobic semiconductor.\cite{porrazzo_improving_2014,zhang_electrochemical_2020}

Here the difference between an increase in trap density and trap filling needs to be noted. The trap density is described by the trap density-of-states function (trap DOS) and is usually modeled as a gaussian or exponential tail extending into the bandgap.\cite{vissenberg_theory_1998,kalb_calculating_2010} Trap states, i.e. localized states inside the bandgap, are intrinsically present in all organic semiconductors, but the trap density can be further increased by exposure to light, moisture or oxygen, by electrical stress or by a combination of these.\cite{gomes_electrical_2005,sirringhaus_reliability_2009,bobbert_operational_2012,haneef_charge_2020,kalb_oxygen-related_2008,cielecki_photo-induced_2020,iqbal_suppressing_2021,nicolai_unification_2012,zuo_general_2019} Trap states closer to the band edge are considered shallow, with trap states further inside the bandgap being considered deep traps. 

Trap states can be either permanent or transient irrespective of their depth, as their lifetime depends not on the depth of the state inside the bandgap but on the formation energy of the defect acting as a trap state.\cite{iqbal_suppressing_2021} Trap filling occurs in an OFET as the applied gate voltage moves the charge carrier quasi-Fermi level from deep inside the bandgap towards the band edge. Upon removal of the gate voltage, a trapped charge will be released after a period determined by the depth of the trap. During a transfer measurement, the gate voltage sweep rate needs to be slow enough that trap filling and release has time to occur, reaching a quasi-equilibrium.\cite{kalb_calculating_2010,chen_intrinsic_2018} A too fast sweep rate will lead to an increase in hysteresis, as charges being trapped (released) will retract from (add to) the source-drain current during the on (off) sweep. In this work, trapping and de-trapping times are shown to be short, and thus we only concern ourselves with changes in the trap density in the analysis.

Poly-3-hexylthiophene (P3HT) is a benchmark semiconducting polymer, and its highly hydrophobic nature lends it well to water-gated applications. P3HT WG-TFTs have demonstrated good environmental stability, with only minor surface changes due to water exposure and a long shelf-life when protected from light and oxygen exposure.\cite{picca_study_2019} In the biosensor configuration, the device is operated with a gate covered by a self-assembled monolayer (SAM). Typically, the SAM contains biorecognition elements selective to the biomolecule of interest. When the specific biomolecule reaches the biorecognition element, there will be a change in the gate potential and/or its capacitance, leading to a measurable change in the source-drain voltage and the device figures of merit (FOM) - mainly the threshold voltage.\cite{wang_electrolytic_2016,berto_egofet_2018,macchia_single-molecule_2018} Care needs to be taken during sensing measurements, as similar current changes can also be caused by changes in the semiconductor bulk or surface, by contamination of the water or by any combination of these. To ensure that the measured current change is due to the biorecognition element, intermittent measurements using a bulk gold gate without surface treatment, henceforth called the reference gate, are performed to ensure all other factors remain stable.\cite{picca_ultimately_2020,macchia_single-molecule_2018} 

In this work, we monitor a P3HT-based WG-TFT over two months and 5750 transfer cycles. Biosensing experiments are performed throughout this time, and the channel stability is continuously probed using reference gate measurements. Assessing the long-term changes in device figures of merit during normal device usage, we show that device degradation is caused mainly by electrical stress and progresses at a constant rate throughout the measurements. The shift in threshold voltage and decrease in mobility throughout the device lifetime is consistent with an increase in trap density, which we also observe using estimations from the subthreshold slope. In addition to a steady increase in trap density, we observe a steep rise in trap density during heavy usage and a partial recovery during rest, indicating the formation of two different trap species.

\section{\label{sec:experiment}Experiment}
\subsection{\label{sec:fabrication}Fabrication of the P3HT WG-TFT}
\begin{figure}[h]
\includegraphics[width=\linewidth]{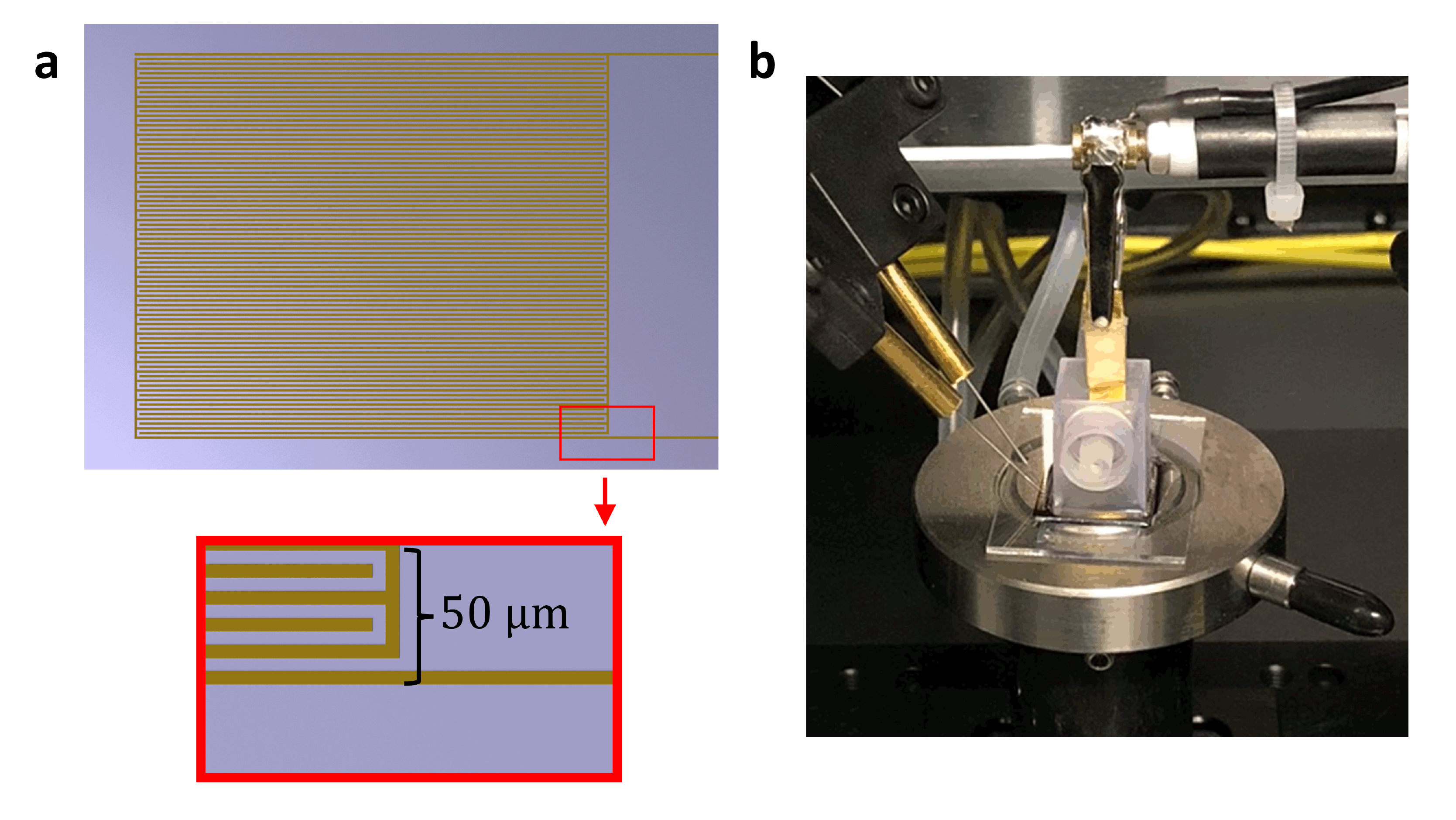}
\caption{\label{fig:2} (a) Drawing of the interdigitated source and drain electrodes with $W/L=16160$. (b) Measurement setup inside the probe station showing source and drain contact probe needles, solid gold reference gate and resistor / well assembly. The external reservoir is not shown in the picture.}
\end{figure}

Devices were manufactured using P3HT acquired from TCI, which at the time had the highest regioregularity commercially available ($RR>99\%$ and $M_n=27000-45000$). P3HT was dissolved in chlorobenzene under ambient conditions at a concentration of $4\,\text{mg}\,\text{ml}^{-1}$. The solution was sonicated for 20 min before filtering through a 0.2\,$\mu$m PFTE filter. Patterned substrates were acquired from the Technical Research Centre of Finland. They consist of silicon covered by a layer of thermally grown $\text{SiO}_2$. A 5\,nm thick chromium adhesion layer followed by 50\,nm of gold forms interdigitated electrodes on top, illustrated in Fig.~\ref{fig:2}a along with the measurement setup in Fig.~\ref{fig:2}b. The channel is 80800\,$\mu$m wide and 5\,$\mu$m long, for a $W/L$ ratio of 16160. Substrates were ultrasonically cleaned in acetone and IPA, 10\,min in each, before rinsing with DI-water. 100\,$\mu$l of the semiconductor solution was deposited using spin coating in ambient air, at 2000\,rpm for 30\,s. Samples were annealed at 90\,°C for 60\,min in darkness. Resistors were stored in darkness in a vacuum before measurements. In preparation for electrical measurements, 3D printed polymer water wells were fitted to the resistor using PDMS. The well can hold approximately 1\,ml and was filled with HPLC-grade water for use as the gate dielectric. It is connected to a larger external reservoir to keep the water level constant during measurements. A rectangular piece of gold was used as the reference gate electrode and held vertically with a clamp. 

The studied WG-TFT was one of several used for biosensing experiments. It was chosen for analysis due to mostly identical experiments being conducted throughout the device lifetime, offering the most consistency in measurement conditions. 

\subsection{\label{sec:measurement}Electrical measurements}Measurements were conducted under ambient conditions (20\,\% relative humidity at 23\,°C) in the dark. A Keithley 4200A SCS Parameter Analyzer and a probe station were used for all electrical measurements. A drain voltage ($V_{DS}$) of -0.4\,V was applied while sweeping the gate voltage ($V_{GS}$) from +0.1\,V to -0.4\,V and back at a rate of 20\,mV\,$\text{s}^{-1}$.
\section{\label{sec:results}Results \& Discussion}
\subsection{\label{sec:stability}Long-term stability analysis of a single WG-TFT}
\begin{figure*}[h]
\includegraphics[width=\linewidth]{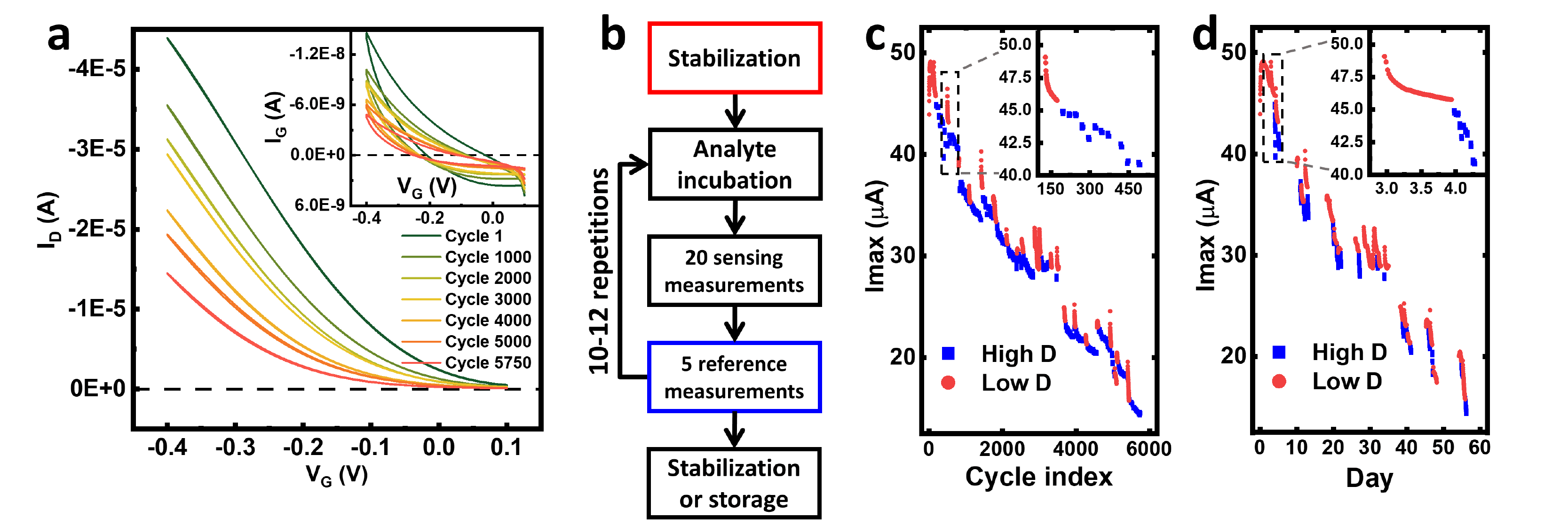}
\caption{\label{fig:3}(a) Evolution of the transfer and gate current over 5750 cycles from two months of biosensing measurements. The measurement protocol (b) describes a typical session of sensing measurements. The stabilization and reference measurements are performed with the same solid gold reference gate throughout the device lifetime, while sensing is performed with a single-session use gate with a functionalized gold surface. The maximum drain current measured during each stabilization (low duty, see Eq.~(\ref{eg:one})) cycle and reference cycle (high duty) is presented in (c) and (d) as a function of cycle index and time, respectively. The insets show the first sensing measurement session after the initial stabilization of the device}
\end{figure*}

Fig.~\ref{fig:3}a shows the change in device transfer characteristics over two months and 5750 transfer measurement cycles, with the inset showing the corresponding gate currents. The maximum S-D current decreases from around 45\,$\mu$A to 15\,$\mu$A during this time. The hysteresis in the transfer curve remains negligible throughout the measurements, and the source-drain current is three orders of magnitude larger than the gate current, both of which indicate a device operating in a proper voltage region without incurring any significant electrochemical degradation.\cite{picca_ultimately_2020} The lack of hysteresis also indicates that trapping and de-trapping times are significantly shorter than the transfer measurement time, meaning the measurements take place in a quasi-equilibrium.\cite{kalb_calculating_2010}
The measurement protocol for sensing measurements is depicted in Fig.~\ref{fig:3}b. More in detail, the device is stabilized using the bulk gold gate and a long delay between measurement cycles until the change in current is below 2\,\% per hour. After initial stabilization has been achieved, a functionalized gate is inserted into the water well next to the reference gate. The reference gate is electrically disconnected and left in the well. Then, 20 transfer cycles are recorded with a delay of 30 seconds between each measurement using the functionalized gate. The functionalized gate is subsequently removed from the device and incubated in the analyte for 10\,min, during which time the electrical connection to the reference gate is restored and five transfer cycles are recorded with a delay of 30 seconds. This way, the stability of the channel and electrolyte can be continuously probed, ensuring that any significant change observed in device characteristics during sensing measurements is caused purely by changes to the surface of the functionalized gate. This method also enables the analysis of long-term changes in device FOM such as the threshold voltage, hole mobility, and effective trap density, as the same gold gate is used for stabilization and reference measurements throughout the device’s lifetime.
Here we introduce a helpful metric to investigate the effect of electrical stress on device degradation. The duty cycle, D, is the fraction of time an electrical bias is applied and is defined as
\begin{equation}
\label{eg:one}
    D = \frac{\text{time under bias}}{\text{time under bias + time resting}}
\end{equation}
The measurements can be divided into low and high duty cycle measurements (low $D$ and high $D$). The duration of a single transfer measurement is 46 seconds. During stabilization, the delay time is 30\,min and gives low $D$ $\approx3$\,\%. During sensing and reference measurements, the delay time is 30\,s which gives a high duty cycle of $D$ $\approx61$\,\%.

Fig.~\ref{fig:3}c shows the evolution of the maximum drain current over 5750 cycles plotted against the cycle index. Note that cycles measured during sensing measurements are also included in the cycle index. The current decreases nearly linearly as a function of cycle index over the device lifetime. When the same measurements are plotted as a function of time (Fig.~\ref{fig:3}d), the maximum drain current rapidly changes as the duty cycle is changed from the low stress of the stabilization to high stress during measurements. The insets in Fig.~\ref{fig:3}c and ~\ref{fig:3}d show the change in maximum drain current over the first 450 cycles of operation following the initial stabilization. Assuming device degradation results from P3HT exposure to water, it would be expected to have a time-dependent trend, but this is not observed in the measurements. Conversely, when the current is plotted as a function of cycle number, the change in duty cycle does not significantly change the shape of the curve, indicating that the change is driven by electrical stress. The same pattern is repeated throughout the whole cycle index range.
\begin{figure*}
    \includegraphics[width = \linewidth]{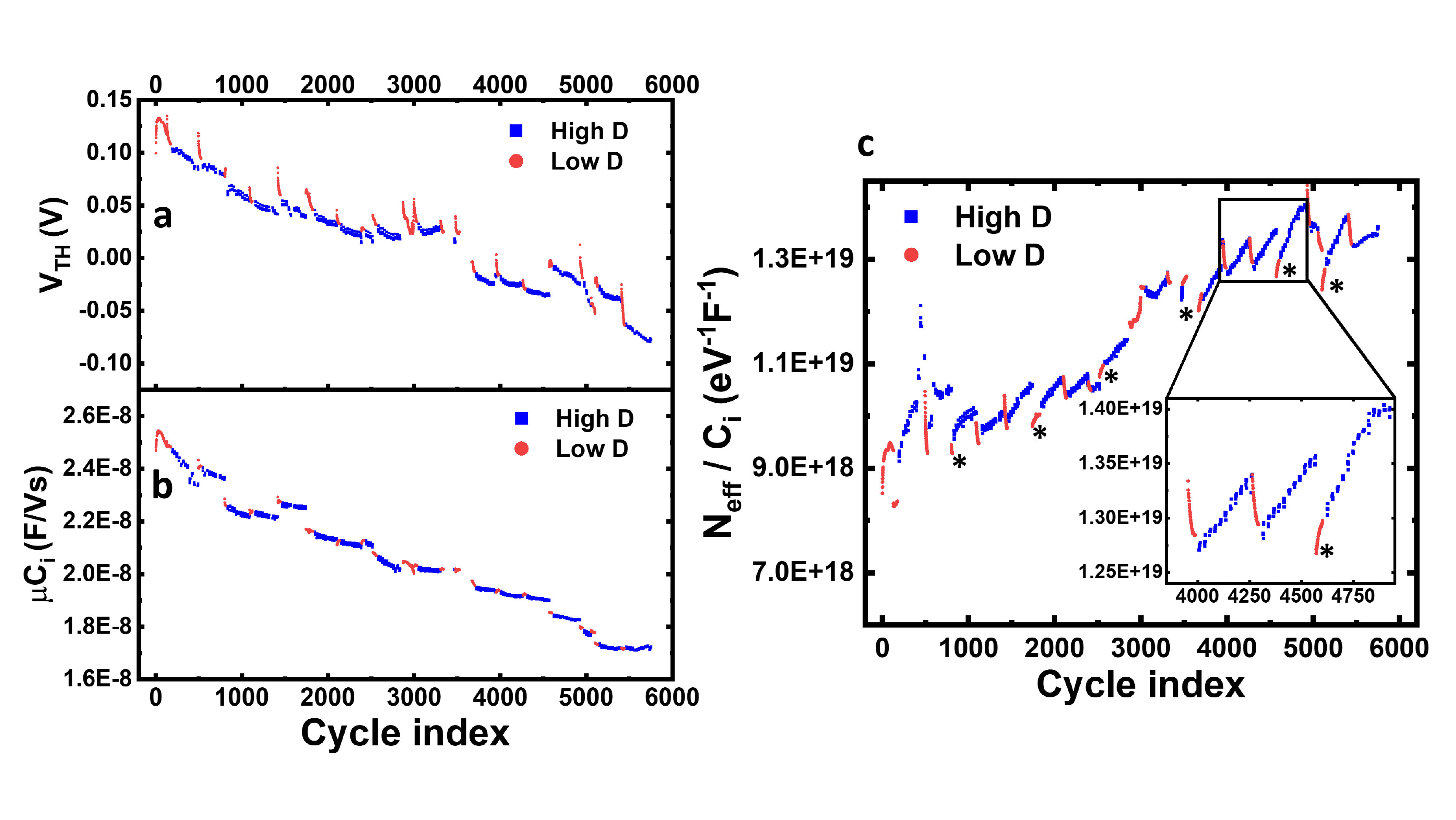}
    \caption{(a) Threshold voltage $V_{TH}$  as a function of cycle index, showing a nearly linear shift over the long term. The effect of gate exposure to air manifests as faster shifting during some low $D$ measurements. (b) $C_i\mu$ as a function of cycle index showing a nearly linear shift but no difference between low and high $D$ measurements. (c) $[N_{eff} \/C_i]$ as a function of cycle index. The stars indicate measurements conducted after the device has been resting for at least two days. The inset highlights an example of net increase in trap states during high $D$ and a net decrease during the following low $D$ measurements.}
\label{fig:4}
\end{figure*}
The source-drain current in the saturation region - in which the device is operated - is given by the following equation,
\begin{equation}
\label{eq:two}
    I_D = \frac{W}{L}\frac{C_i \mu }{2}(V_G-V_{TH})^2
\end{equation}
where W and L are the width and length of the channel. The channel capacitance in similar devices has been measured to be on the order of  3\,-\,10\,$\mu$F\,$\text{cm}^2$.\cite{kergoat_water-gate_2010,cramer_double_2012,melzer_characterization_2014} Using $C_i=5$\,$\mu$F\,$\text{cm}^2$, we arrive at a mobility of around $5\times10^{-3}$\,$\text{cm}^2\,\text{V}^{-1}\,\text{s}^{-1}$ for the fresh transistor, in line with earlier reported values.\cite{melzer_characterization_2014,salleo_intrinsic_2004,kline_dependence_2005} However, the channel capacitance $C_i$ and the mobility $\mu$ cannot be decoupled utilizing only transfer measurements and therefore the product $[C_i\mu]$ will instead be considered. The threshold voltage, $V_{TH}$, and $[C_i\mu]$ were determined from the $\sqrt{I_d}$ vs $V_G$ plot for each cycle using a python script and are shown in Fig.~\ref{fig:4}a and ~\ref{fig:4}b, respectively. The Python script is described in more detail in Supplementary information section S1. The subthreshold slope $S$ can be extracted from a linear fit to log$I_D$ vs $V_G$ in the subthreshold region and provides information on the effective trap density according to
\begin{equation}
\label{eq:three}
    S=\frac{k_B T \text{ln}10}{q}\left[1+\frac{N_{eff}}{C_i}q^2\right]
\end{equation}
where $N_{eff}$ is the effective trap density per unit energy and unit area, $k_B$ is the Boltzmann constant, $T$ is the temperature, $q$is the elementary charge.\cite{kalb_calculating_2010,kalb_oxygen-related_2008,kwon_subthreshold_2016} Using literature values for the capacitance, the trap density is on the order of $5\times10^{13}$\,$\text{eV}^{-1}\,\text{cm}^{-2}$, which is on the higher side of earlier reported values using the method.\cite{kalb_calculating_2010,mcdowell_improved_2006} It is not possible to decouple the effect of any change in the capacitive coupling between the gate, electrolyte, or electrolyte-semiconductor surface from that of an increase in trap density, so we consider $[N_{eff}\/C_i]$ instead. The evolution of $[N_{eff}\/C_i]$ as a function of cycle index is shown in Fig.~\ref{fig:4}c. The trap DOS can be determined in conventional OFETs utilizing temperature series measurements, but this is not feasible with a WG-TFT. The effective trap density used here assumes a constant trap DOS and can be considered a rough but useful estimate.\cite{kalb_calculating_2010}

The trap density is represented by $[N_{eff}\/C_i]$ in Fig.~\ref{fig:4}c and is indeed observed to increase. The increase during high $D$ is linear with cycle index and very consistent throughout the device lifetime. This indicates gate biasing as a driver of the increasing trap density, as opposed to source-drain current; the S-D current decreases significantly towards later cycles, but the rate of increasing trap density remains the same. 

During low $D$, a net decrease in trap density is observed when high $D$ measurements have been performed beforehand. However, when measurements are initiated after a rest period, marked with * in Fig.~\ref{fig:4}4c, the trap density increases similarly to that during high $D$ measurements. This can be explained as generation of two different trap species during operation. Shorter-lived trap states will disappear during rest periods while longer-lived ones will remain, leading to the long-term increase in trap density observed in Fig.~\ref{fig:4}c as well as the partial recovery observed during low $D$ and rest. Generation of such metastable traps in OFETs has previously been reported by Iqbal, et al. and attributed to the formation of radical complexes involving water.\cite{iqbal_suppressing_2021} By fitting an exponential decay to the instances of net de-trapping under low $D$, the lifetime of the shorter-lived trap species was estimated to be $437\pm225$ minutes.

As we found a decrease in trap density during low $D$ measurements, and an increase during high $D$ measurements, there would be some delay time at which the change in trap density would be minimal. By fitting the data displayed in the inset of Fig.~\ref{fig:4}c with linear fits, assuming a bias-induced increase and time-dependent decrease in trap density, the ideal delay time for low $D$ measurements was found to be around 5 minutes. Using this delay time, the trap density should remain stable during low $D$ measurements, shortening the stabilization time needed when switching to high $D$ measurements. This could improve device stability during long measurements and improve consistency between measurement days. However, the ideal delay time may vary between devices and geometries.
\begin{figure}
    \centering
    \includegraphics[width=\linewidth]{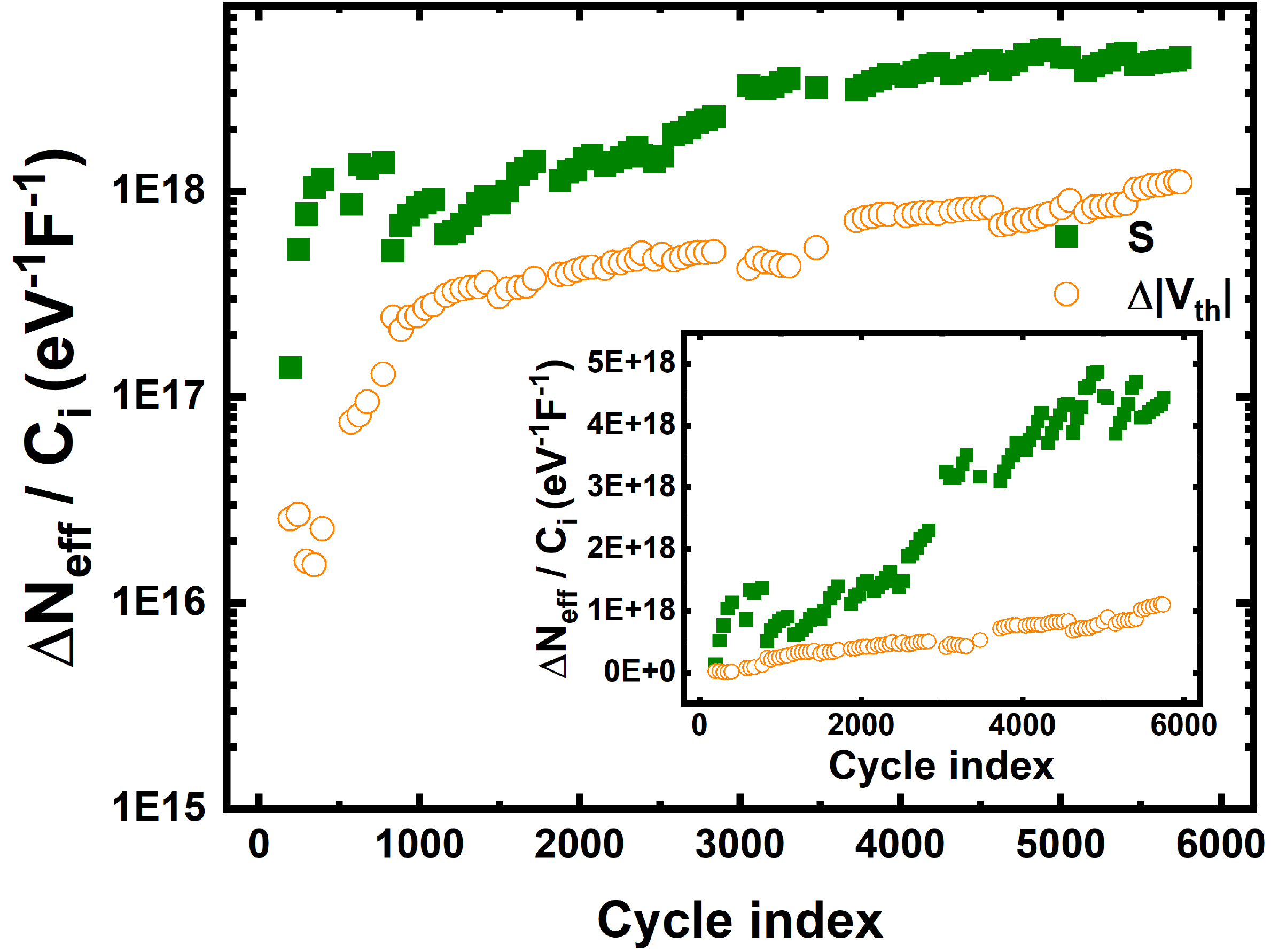}
    \caption{Change in $[N_eff\/C_i]$ calculated using the $S$ and $\Delta V_{TH}$ methods. The inset shows the plot on a linear scale. Only data from high duty cycle measurements are shown for clarity.}
    \label{fig:5}
\end{figure}
The increase in trap density can also be approximated utilizing the shift in threshold voltage:\cite{kalb_oxygen-related_2008,lamport_tutorial_2018}
\begin{equation}
\label{eq:four}
    N_{eff} \approx \frac{C_i |{\Delta V_{TH}}|}{q}
\end{equation}
Fig.~\ref{fig:5} shows a comparison of the increase in $[N_{eff} \/C_i]$ calculated using the $S$ and $\Delta V_{TH}$ methods in semi-log and linear scale (inset). For clarity, only high $D$ data is shown. The trap density calculated using the $S$ method increases faster throughout the device’s lifetime. As the quasi-Fermi level is located further from the band edges in the subthreshold region, the $S$ method probes deeper trap states than the $\Delta V_{TH}$ method.\cite{haneef_charge_2020,kalb_oxygen-related_2008} The more significant increase observed with the $S$ method could then be understood as deeper traps dominating trap state generation, in agreement with earlier studies.\cite{iqbal_suppressing_2021,podzorov_intrinsic_2004} However, the double layer capacitance at the semiconductor surface will increase with applied gate voltage, limiting the usefulness of the comparison as the values are calculated at different gate voltages.\cite{macchia_single-molecule_2018} 

\subsection{\label{sec:Gate exposure to air}Short-term shift in the threshold voltage upon exposing the gate to air}
\begin{figure}
    \centering
    \includegraphics[width=\linewidth]{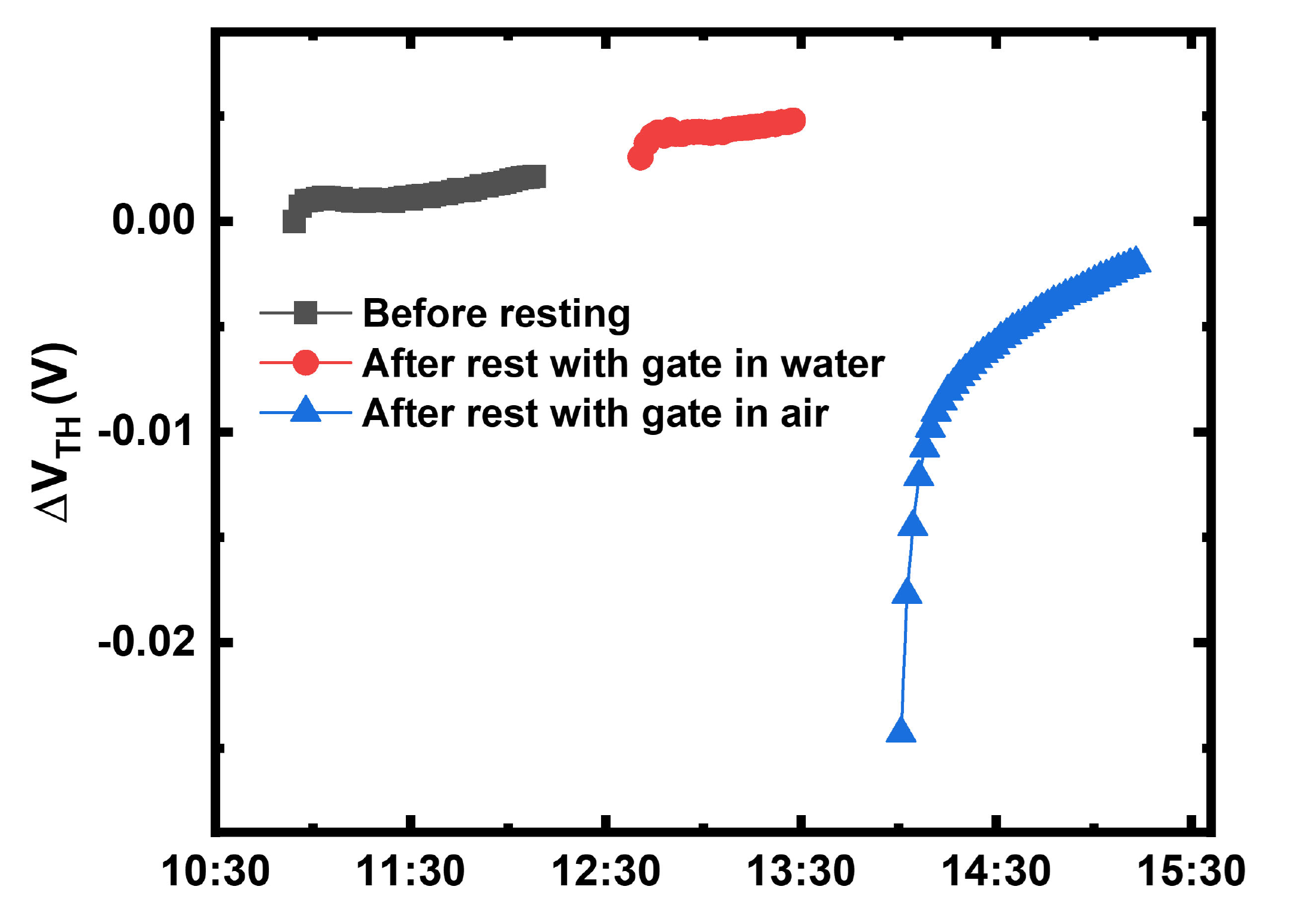}
    \caption{Gate voltage shift observed after letting the device rest 15 minutes with the gate electrode either submerged in the well or exposed to air. After exposing the gate to air, the device exhibits a significant threshold voltage shift before returning to the preceding level. }
    \label{fig:6}
\end{figure}
A brief investigation of gate exposure to air was conducted on an identically prepared WG-TFT. The device underwent initial stabilization by cycling with a delay of 30 minutes over two days before the measurement was started. Transfer measurements were conducted before and after resting the device for 15 minutes with the gate either submerged in the well or removed and kept in air. The shift in threshold voltage as a function of time is shown in Fig.~\ref{fig:6}. After resting the device with the gate submerged in water, a negligible shift in threshold voltage can be observed, with the device remaining stable afterward. However, after the gate was exposed to air, a larger shift in threshold voltage can be observed, with a subsequent return towards the initial level. This behavior is like that observed in the previously used device when resuming measurements after storage. During storage, the gate is kept submerged in water to suppress this effect, however during reassembly of the device the gate is dried and thus exposed to air before reinsertion into the WG-TFT well. The exact mechanism of the gate potential shift upon exposure to air warrants more research but can likely be attributed to adsorption or (partial) oxidation giving rise to a shift in work function.\cite{wells_adsorption_1972}

\section{\label{sec:conclusions}Conclusions}
The long-term stability of a P3HT WG-TFT was analyzed over two months, during which the device was in intermittent usage, following a protocol for biosensing measurements. It was shown that the change in device figures of merit was caused by electrical stress, with time spent submerged in water having a negligible effect on the device. We show that hysteresis and gate leakage remained low throughout the measurements, with degradation manifesting as a decrease in overall current. The decrease in current was attributed to a linear shift in the threshold voltage and a linear decrease in mobility with cycle number, both of which are consistent with an increase in trap density in the semiconductor.
The trap density was approximated using the subthreshold slope, showing that a net increase occurred during periods of high duty cycle measurements, with a decrease occurring during low duty cycle measurements and rest. Generated trap states consisted of a near-permanent species and a species with a lifetime of several hours. The long usable lifetime utilizing a simple geometry and commercially available semiconductor further support WG-TFTs in biosensing applications. Our investigations provide new insights into the degradation mechanisms dominating the device’s short- and long-term behavior. However, more specialized long-term measurements in conjugation with robust modeling are still needed to decouple the effects of the different device parts and fully understand the stability issues.

\section{\label{sec:acknowledge}Acknowledgments}
Financial support from the Academy of Finland through projects $\#316881$,  $\#316883$, and $\#270010$; H2020 – Electronic Smart Systems – SiMBiT: Single-molecule bio-electronic smart system array for clinical testing (Grant agreement ID: 824946), and \r{A}bo Akademi University CoE "Bioelectronic activation of cell functions" are acknowledged.

\nocite{*}

\bibliography{references}

\end{document}